\documentclass[aps,preprint,prd,showpacs]{revtex4}

\usepackage{amsmath}
\usepackage{graphicx}
\usepackage{subfigure}
\usepackage{multirow}
\usepackage{bm}
\usepackage{amssymb}
\usepackage{mathtools}
\usepackage{enumerate}
\usepackage{color}
\usepackage[colorlinks,linkcolor=magenta,anchorcolor=cyan,citecolor=blue]{hyperref}
\usepackage{soul}

\begin{document}
    \title{Shape of CMB lensing in the early dark energy cosmology}
    \author{Gen Ye$^{1}$\footnote{ye@lorentz.leidenuniv.nl}}
    \author{Jun-Qian Jiang$^{2}$\footnote{jiangjq2000@gmail.com}}
\author{Yun-Song Piao$ ^{2,3,4,5} $\footnote{yspiao@ucas.ac.cn}}

 \affiliation{$^1$ Institute Lorentz, Leiden University, P.O. Box 9506, Leiden 2300 RA, The Netherlands}
 \affiliation{$^2$ School of Physical Sciences, University of Chinese Academy of Sciences, Beijing 100049, China}

\affiliation{$^3$ School of Fundamental Physics and Mathematical
    Sciences, Hangzhou Institute for Advanced Study, UCAS, Hangzhou
    310024, China}

    \affiliation{$^4$ International Center for Theoretical Physics
    Asia-Pacific, Beijing/Hangzhou, China}

\affiliation{$^5$ Institute of Theoretical Physics, Chinese
    Academy of Sciences, P.O. Box 2735, Beijing 100190, China}

    \begin{abstract}

Recently, the cosmological tensions, $H_0$ and $S_8$ in
particular, have inspired modification of both pre- and postrecombination physics simultaneously. Early dark energy is a
promising pre-recombination solution of the $H_0$ tension, known to be compatible
with the cosmic microwave background (CMB). However, the compatibility of early dark energy, as well
as general early resolutions, with the CMB is no longer obvious if the
late Universe is also modified. Aside from cosmological
parameters, the main channel through which late Universe physics
affects CMB observables is gravitational lensing. We employed a new method of sampling functions using the Gaussian Process in the Monte Carlo Markov Chain analysis to constrain the shape of the CMB lensing potential. We obtained the early Universe (CMB) only constraints on the full shape of the CMB lensing
potential, with the late-time Universe being marginalized over. It is found that CMB data prefers a lensing potential shape
that is $\Lambda$CDM-like at $80\lesssim L\lesssim400$ but with enhanced
amplitude beyond this range. The obtained shape constraints can
serve as a CMB-compatibility guideline for both late and early
Universe model building that modifies the lensing potential.

    \end{abstract}
    \maketitle
    \section{Introduction}\label{sec:intro}
    One of the most important parameters in cosmology is the Hubble constant $H_0$, which characterizes the expansion rate of our current Universe. It can be measured directly through the distance-redshift relation. Using the state-of-art Cepheid calibrated distance ladder approach, the SH0ES group reports $H_0=73.04\pm1.04$ km/s/Mpc \cite{Riess:2021jrx}. On the other hand, in a given cosmological model $H_0$ calibrates the distance to the last scattering surface and can be derived from precise cosmic microwave background (CMB) measurements. The Planck satellite results yield $H_0=67.37\pm0.54$ km/s/Mpc assuming $\Lambda$CDM \cite{Planck:2018vyg}, also confirmed by more recent ground CMB experiments \cite{ACT:2020gnv, SPT-3G:2021wgf}. The $>4\sigma$ tension, usually dubbed the Hubble tension, between the locally measured $H_0$ and the CMB derived ones is one of the most significant challenges faced by current cosmology, which might be a hint of new physics beyond $\Lambda$CDM \cite{Verde:2019ivm,Shah:2021onj}.

Among the plethora of models proposed to alleviate the Hubble
tension, see e.g. \cite{DiValentino:2021izs,
Perivolaropoulos:2021jda} for thorough reviews, a promising class
of solutions is to introduce beyond $\Lambda$CDM modifications to
the pre and near recombination evolution (early resolution), such
as early dark energy (EDE)
\cite{Poulin:2018cxd,Agrawal:2019lmo,Alexander:2019rsc,Lin:2019qug,Smith:2019ihp,Niedermann:2019olb,Sakstein:2019fmf,Ye:2020btb,Braglia:2020bym,Karwal:2021vpk,Niedermann:2021vgd,McDonough:2021pdg,McDonough:2022pku,Rezazadeh:2022lsf, Brissenden:2023yko, Seto:2021xua},
see e.g. \cite{Kamionkowski:2022pkx,Poulin:2023lkg} for recent
reviews, early modified gravity \cite{Lin:2018nxe,
Zumalacarregui:2020cjh, Braglia:2020iik, Ballesteros:2020sik},
neutrino self-interaction \cite{Blinov:2019gcj, Kreisch:2019yzn},
nonstandard recombination process \cite{Chiang:2018xpn,
Jedamzik:2020krr, Sekiguchi:2020teg, Fung:2021wbz} and modified
physical constants at early times \cite{Hart:2019dxi}.

On the other hand, there is another possible tension between large scale
structure (LSS) and CMB observations, characterized by $S_8=\sigma_8
(\Omega_m/0.3)^{0.5}$, where $\sigma_8$ is the root-mean-square
amplitude of the matter fluctuation in spheres of radius
$8h^{-1}$Mpc today. Planck CMB measurement reports $S_8=0.834 \pm
0.016$ with $\Lambda$CDM \cite{Planck:2018vyg} while weak lensing
measurements find lower values. The 3$\times$2pt analysis from
KiDS-1000 obtains $S_8=0.766^{+0.020}_{-0.014}$
\cite{Heymans:2020gsg} and DES-Y3 obtains
$0.779^{+0.014}_{-0.015}$ \cite{DES:2021wwk}. The most recent DES-Y3+KiDS-1000 result, however, gives $S_8=0.790^{+0.018}_{-0.014}$ which is compatible with Planck at 1.7$\sigma$ \cite{Kilo-DegreeSurvey:2023gfr}. Many models that
modify the evolution of the late Universe have been proposed to
resolve this tension, see e.g. \cite{Perivolaropoulos:2021jda,
Abdalla:2022yfr} for recent reviews.

The $S_8$ tension not only affects the $\Lambda$CDM model, but also becomes a concern for EDE \cite{Hill:2020osr,Ivanov:2020ril,DAmico:2020ods}. This tension is usually exacerbated in these early resolutions as a consequence of increased total matter energy density \cite{Ye:2020oix,Pogosian:2020ded,Jedamzik:2020zmd,Vagnozzi:2021gjh}.
Recently there has been increasing interest in reconciling EDE
with LSS by paring it with other possible mechanisms to alleviate the $S_8$ tension, many of which are non-negligible in the post-recombination Universe, such
as axion dark matter \cite{Allali:2021azp,Ye:2021iwa}, decaying
dark matter \cite{Clark:2021hlo}, neutrino mass
\cite{Gomez-Valent:2022bku,Reeves:2022aoi} and dynamics of the
dark sector \cite{McDonough:2021pdg,Wang:2022jpo,
Alexander:2022own,Buen-Abad:2022kgf,Gomez-Valent:2022bku,Wang:2022nap,Reboucas:2023rjm}.
An immediate question follows is what information about EDE we
could extract from the early Universe observations, i.e. CMB,
independent of possible modification to the evolution of the
Universe after recombination. Inversely, such constraints provide
guidelines on how late modification should behave to maintain
compatibility with CMB observations.

    \begin{figure}
   \includegraphics[width=0.9\linewidth]{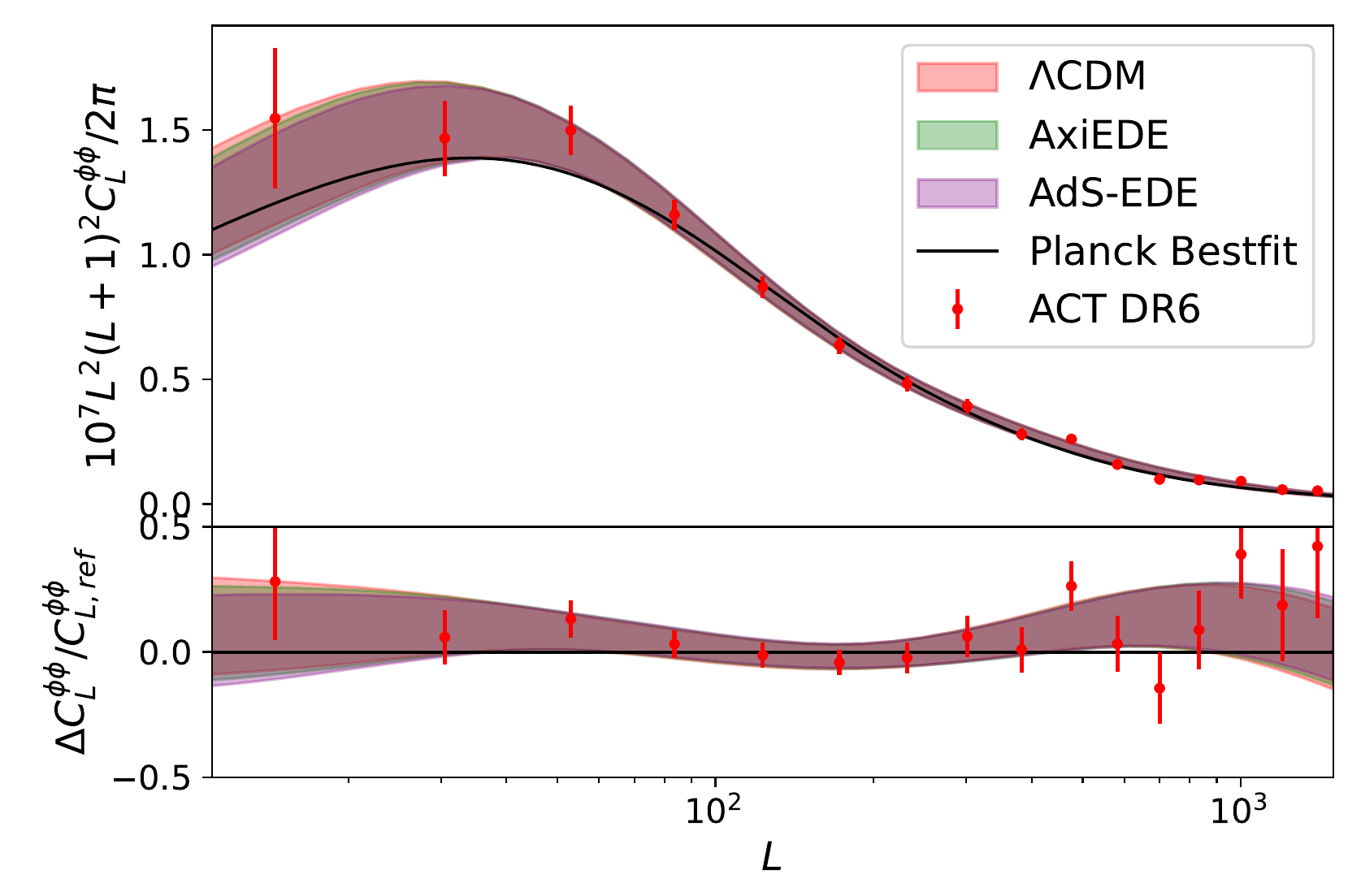}
       \caption{Constraints on lensing spectra shape using the Planck CMB temperature and polarization data as well as Planck + ACT DR6 lensing measurement in the $\Lambda$CDM and EDE models. Shaded areas represent the 95\% distribution of the lensing spectra. Solid black lines are the reference $\Lambda$CDM model from Planck2018 baseline best fit \cite{Planck:2018vyg}. Points with error bars are ACT DR6 \cite{ACT:2023dou} lensing bandpower in the extended multiple range. Following Ref.\cite{ACT:2023dou}, only lensing data in the multiple range $40<L<1300$ is used in the shape constraint, thus the left most two ACT data points do not contribute to the shape constraints. The derived shape constraints contain both information from the Planck+ACT lensing potential reconstruction and the lensed CMB temperature and polarization spectra.}
       \label{fig:clkk_main}
    \end{figure}

On the background level, many late time modifications can be marginalized by a time dependent dark fluid equation of state $w(z)$ and it has been shown in both $\Lambda$CDM \cite{Planck:2018vyg} and EDE \cite{Wang:2022jpo,Reboucas:2023rjm} that a $\Lambda$CDM-like, i.e. $w\simeq-1$, post-recombination Universe is preferred. On the perturbation level, late time physics imprints itself in the CMB observation mainly through the late integrated Sachs-Wolfe (ISW) effect, scattering of CMB photon by the free electrons from reionization and gravitational lensing. It has been suggested that the CMB lensing information can be marginalized by e.g. parametrization \cite{Audren:2012wb, Audren:2013nwa, Motloch:2016zsl, Motloch:2017rlk,Motloch:2018pjy, Motloch:2019gux} and interpolation \cite{Lemos:2023xhs} in $\Lambda$CDM.
In this paper, we employ a new method of sampling free functions in the Monte Carlo Markov Chain (MCMC) analysis. We draw our
functions directly from a Gaussian process (GP) which is able to
explore a wide range of function shapes while also requiring fewer
parameters than the interpolation method. Using the new method, we build the CMB-only constraints on the shape of
lensing in the EDE models, marginalizing over the late Universe evolution,
see Fig.\ref{fig:clkk_main}. We found that CMB data
is most constraining on the shape of lensing in the multiple range
$80\lesssim L\lesssim400$, where the constraints is compatible with $\Lambda$CDM prediction.
Beyond this region, shape constraints deviate from $\Lambda$CDM prediction
and seem to prefer higher amplitudes of the lensing potential spectrum on
both low and high $L$ parts, related to the possible preference
for excess lensing-like smoothing in Planck CMB spectra (lensing
anomaly) \cite{Planck:2018vyg}. Due to the lensing B-mode, CMB B-mode
data from BICEP/Keck \cite{BICEP:2021xfz} contributes nontrivial
constraining power on the $L>400$ tail of the lensing spectrum,
see also Ref.\cite{Ye:2022afu}. Our results show that inclusion of
the B-mode does not remove the preference for a higher lensing
potential on the high-$L$ tail, see Fig.\ref{fig:clkk_bk18}.

    In section-\ref{sec:method} we describe our sampling method and datasets. Section-\ref{sec:results} reports the cosmological models studied and the corresponding results. We conclude in section-\ref{sec:conclusion} and group large plots and auxiliary details in the appendices.

    \section{Methods and datasets }\label{sec:method}
    In this paper we are interested mainly in the linear multiple range $L<3000$ and reionization effects can be well captured by an effective optical depth $\tau_{reion}$. In the high-$l$ CMB data, there is a degeneracy between $\tau_{reion}$ and the primordial scalar power spectrum amplitude $A_s$ as $A_se^{-2\tau_{reion}}$. This degeneracy can be broken by the low-$l$ E polarization data as well as lensing \cite{Galli:2014kla}. Reionization happens at very high redshift $z\gtrsim6.5$ \cite{Fan:2006dp}, on which late time dynamics relevant to LSS generally has marginal effect. Therefore, aside from cosmological parameters, the main interface between the free streaming CMB photons after recombination and possible post-recombination modifications relevant to LSS is the late ISW effect and gravitational lensing. The late ISW effect receives constraints from low-$l$ TT spectrum where data is plagued by cosmic variance. Appendix-\ref{apdx:isw} shows that it has marginal constraining power on details of the late-time physics by marginalizing over the overall strength $A_{LISW}$ of the late ISW effect. Therefore, we assume standard late ISW contribution ($A_{LISW}=1$) and only focus on the CMB lensing spectrum in this paper.

    Instead of the theory lensing power spectrum $C^{\phi\phi}_{L,\text{th}}$ predicted by a model that describes the full post-recombination evolution, we use
    \begin{equation}\label{eq:cl}
        C^{\phi\phi}_{L}=A(L) C^{\phi\phi}_{L,\text{th}}
    \end{equation}
     to calculate the lensed temperature and polarization power spectra and fit to CMB data, where $A(L)$ is a free function to be sampled from a GP, which can be thought of as a Gaussian distribution of functions $f(x) \sim GP\left[\bar{f}(x), K\right]$ with $\bar{f}$ the mean function and $K$ the kernel function (``covariant matrix"). We use the standard square exponential kernel $K(x_1, x_2) = \sigma^2\exp(-\frac{|x_1-x_2|^2}{2l^2})$, with correlation length $l$ characterizing the smoothness of the function samples and $\sigma$ being the uncertainty amplitude parameter with respect to the mean.

     We use the Monte Carlo Markov Chain (MCMC) method to sample the parameter space. One problem is that the GP sampling process is purely random and receives no ``guidance" from any MCMC likelihoods (data) thus will severely affect convergence speed. To remedy this, we introduce a few nodes $\{\mathsf{X}_i, \mathsf{A}_i\}_{i\in I}$ where $\mathsf{X}_i\equiv\log_{10}\mathsf{L}_i$ are fixed multiple positions in log space and $\mathsf{A}_i = A(\mathsf{L}_i)$ are amplitudes at $\mathsf{L}_i$, which are varied in the MCMC analysis. $A(L)$ is then drawn from a conditional GP under the condition that the sampled function passes through these specified nodes. Formally, on any given finite set of multiples $\{L\}$ we define $X_L=\{\log_{10} L\}$, then the GP reduces to a conditional multivariate Gaussian distribution
    \begin{equation}\label{eq:al}
        A(L) \sim \mathcal{N}\left(\bar{A}(L), \mathbf{\Sigma} \right)
    \end{equation}
    with
    \begin{eqnarray}
        \bar{A}(L) &=& K(X_L, \mathsf{X}_i)K^{-1}(\mathsf{X}_i,\mathsf{X}_j)\mathsf{A}_j\\
        \Sigma_{L_1 L_2} &=& K(X_{L_1}, X_{L_2})-K(X_{L_1}, \mathsf{X}_i)K^{-1}(\mathsf{X}_i,\mathsf{X}_j)K(\mathsf{X}_j, X_{L_2})
    \end{eqnarray}
    where common subscripts are summed over. On one hand, nodes do not bias our function sampling because the amplitudes $\mathsf{A}_i$ are free to vary in the MCMC analysis. On the other hand, nodes aid convergence by shaping the mean function to the data-constrained direction. Moreover, unlike the standard interpolation method, GP still provides sufficient freedom for $A(L)$ to vary even one fixes the nodes and amplitudes, which allows us to use much fewer nodes than standard interpolation. In principle, a large number of Gaussian realizations of $A(L)$ is needed for the same cosmology to sample the function space as much as possible. On the other hand, to perform the MCMC analysis over cosmological parameters, we cannot afford too much time at each cosmological step (i.e. the same cosmological parameters). As a compromise, in one cosmological step we oversample 2 sets of node amplitudes $\{\mathsf{A}_i\}$ and test over 4 random samples of $A(L)$ for each set of $\{\mathsf{A}_i\}$, thus in total 8 realizations of $A(L)$ are sampled in each cosmological step. Each sampled realization of $A(L)$ is treated as a separate MCMC step with its acceptance into the chain depending on the usual MCMC criteria \footnote{The Nth point is rejected if its $\chi^2_{N}$ is larger than the previous step $\chi^2_{N-1}$, otherwise there is a possibility depending on $\Delta\chi^2=\chi^2_{N-1}-\chi^2_N$ of accepting and storing it in the chain.}. Increasing the oversampling number without heavily impacting the convergence speed has little effect on the final constraints. The MCMC sampling is performed using MontePython-3.5 \cite{Audren:2012wb, Brinckmann:2018cvx}, modified to implement the GP sampling method described above \footnote{Available at \url{https://github.com/genye00/montepython-3.5_lens.git}}. Cosmological background and perturbations are calculated using CLASS \cite{Lesgourgues:2011re, Blas:2011rf}, modified to include the EDE models and recomputation of lensing \footnote{Available at \url{https://github.com/genye00/class_multiscf.git}}. The modification exposed the lensing module of CLASS so that it supports redoing the lensing computation only given input vector $A(L), ~L=2,3,\dots,L_{\rm{max}}$ with $A(0)=L_{\rm{max}}$ if cosmology keeps unchanged. For speed reasons, the actual vector drawn from GP is the values of $A(L)$ evaluated at 40 $L$'s log-evenly distributed in the multiple range $10\le L\le1500$, which is then interpolated to the full $A(L)$. The corresponding values of $C_{L,\rm{th}}^{\phi\phi}$ and $A(L)$ at the 40 $L$'s are output into the chain for every step, and posterior distributions of the product \eqref{eq:cl} of them will be later presented as the lensing shape constraint. 

    Our baseline datasets include:
    \begin{itemize}
        \item  \textbf{CMB T/E}: CamSpec High-$\ell$ TTTEEE CMB likelihood \cite{Rosenberg:2022sdy} derived from the Planck PR4 NPIPE data \cite{Planck:2020olo}. Low-$\ell$ CMB TT (PR3) and EE (PR4) data from Planck \cite{Planck:2019nip, Planck:2020olo}.
        \item \textbf{Lensing}: The combined likelihood of ACT DR6 lensing and Planck NPIPE lensing with extended range ($40<L<1300$) \cite{Carron:2022eyg,ACT:2023dou,ACT:2023kun}.
    \end{itemize}
    With explicit mentioning, we will also use the BICEP/Keck 2018 data (BK18) \cite{BICEP:2021xfz} of CMB B-mode polarization and the local measurement of $H_0=73.04\pm1.04$km/s/Mpc \cite{Riess:2021jrx} as a Gaussian prior (R21).

    \section{Models and results}\label{sec:results}
        For $\Lambda$CDM we sample over the standard six cosmological parameters $\{\omega_{cdm}, \omega_b, H_0, \ln10^{10}A_s, n_s, \tau_{reion}\}$. In the typical EDE scenario, a scalar field is initially frozen due to the Hubble friction at some initial position with nonvanishing potential energy. As the Universe cools, the scalar field thaws and starts rolling when the Hubble parameter drops below some critical value and the Hubble friction is no longer able to freeze the field. For specific EDE models we study the $n=3$ axion-like EDE (AxiEDE) \cite{Poulin:2018cxd} and AdS-EDE with a fixed AdS depth $\alpha_{ads}=3.79\times10^{-4}$ \cite{Ye:2020btb} \footnote{According to Ref.\cite{Ye:2020oix}, the existence of an AdS phase is weakly hinted. Here we focus on CMB lensing rather than specific EDE models, thus we choose to fix the AdS depth to the value in Ref.\cite{Ye:2020btb}, which has been shown to be compatible with Planck 2018 CMB spectra and lensing reconstruction.}. EDE is characterized by two additional model parameters $\{z_c, f_{ede}\}$ where $z_c$ is the redshift at which the EDE field starts rolling and $f_{ede}$ its energy fraction at $z_c$. In AxiEDE we vary one more parameter $\Theta_i$ describing the initial phase of the field. See corresponding EDE papers for details. All AxiEDE analysis presented in this paper further includes the R21 $H_0$ prior in addition to our baseline dataset to achieve nontrivial EDE energy fraction \cite{Murgia:2020ryi,Smith:2020rxx}. Without such a prior, the AxiEDE model biases toward $\Lambda$CDM and does not represent the EDE cosmology. AdS-EDE does not suffer this problem partially due to the existence of $V<0$ (AdS), which removes the $\Lambda$CDM ($f_{ede}$) best fit point from this model because too small $f_{ede}$ will
result in the field being confined in the disastrous AdS region and collapse the Universe before today.

We vary amplitudes $\{\ln \mathsf{A}_1, \ln \mathsf{A}_2, \ln
\mathsf{A}_3\}$ with flat priors at three nodes $L\in[50, 200,
800]$ log-evenly distributed in the lensing data multiple
range. Considering that the characteristic shape of the lensing power spectrum $L^2(L+1)^2C^{\phi\phi}_L/2\pi$ is a peak, the three nodes are selected such that the first one near the peak, the second tunes the most constrained amplitude at the intermediate scales and the last anchors the high-$L$ tail \footnote{We have checked that the results will not
significantly depend on the selection of node positions as long as
they are reasonably chosen, i.e. the intervals are large enough and contains most of the data dominated area. Also adding more nodes does not recover finer shape but affects convergence speed negatively.}. Because the typical
relative variance of the lensing data used is $\lesssim 10\%$, we
fix the kernel function parameter to $\sigma=0.1$. We also set the
correlation length parameter $l=1.0$ which is small enough to capture the
main features while also maintains acceptable convergence speed. See Appendix-\ref{apdx:correlation} for details.

\begin{figure}
        \includegraphics[width=0.9\linewidth]{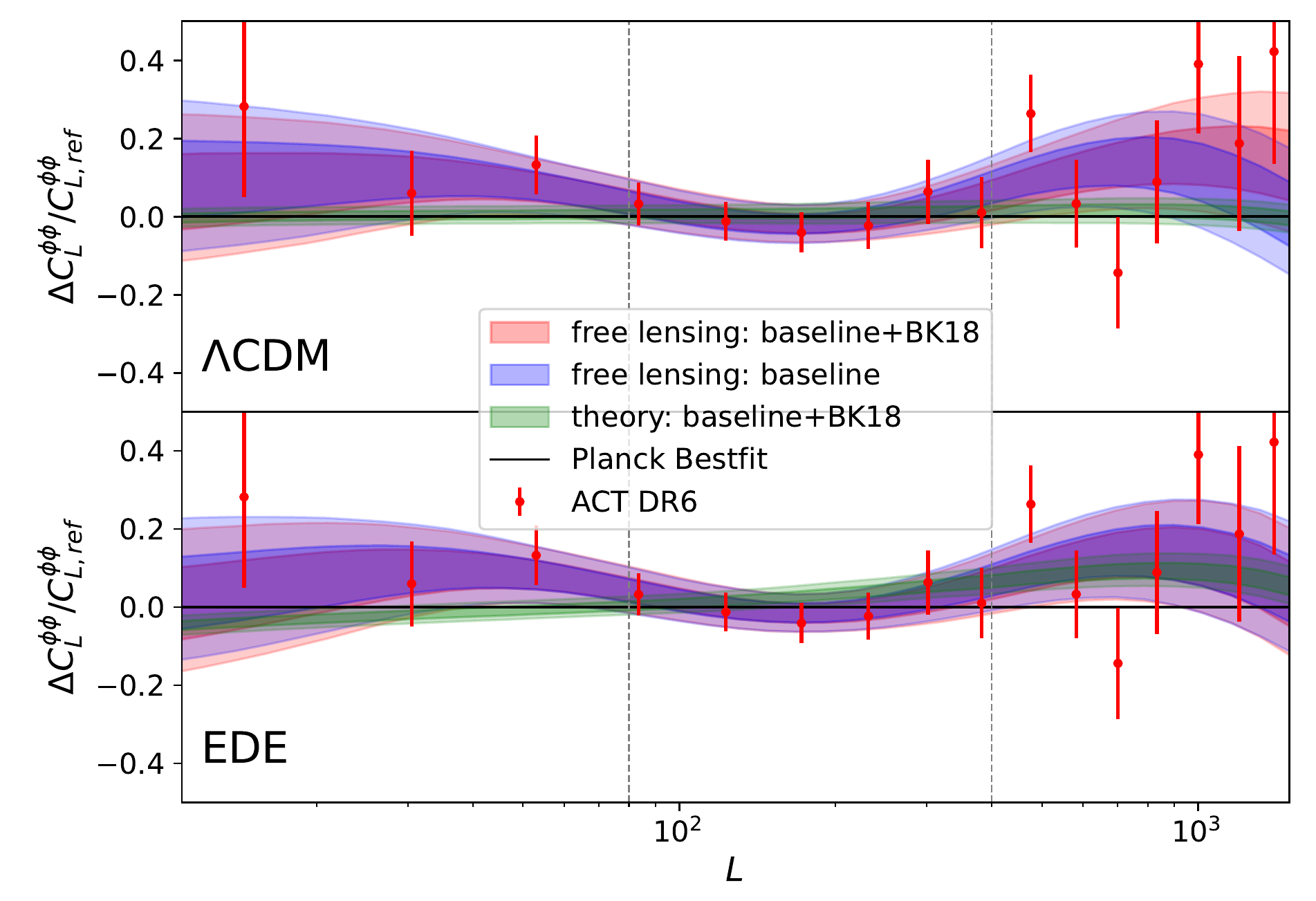}
        \caption{Comparison between the baseline and baseline+BK18 68\% and 95\% constraints on free lensing shape in $\Lambda$CDM and AdS-EDE. The result for AxiEDE is similar to AdS-EDE, thus constraints are only plotted for one EDE model to simplify presentation. The reference model is the  $\Lambda$CDM model from Planck2018 baseline best fit \cite{Planck:2018vyg}. Green constraints are produced by theory derived lensing potential. Dashed gray vertical lines indicate the multiple range $80<L<400$.}
        \label{fig:clkk_bk18}
    \end{figure}

Fig.\ref{fig:clkk_main} plots our main results of the lensing
shape constraints in the baseline dataset for $\Lambda$CDM and
EDE, see Appendix-\ref{apdx:mcmc} for details of the MCMC analysis. Width
of the shaded area at a given multiple indicates the constraining
power of data on lensing potential. One also notes that the constraint are very similar between AxiEDE and AdS-EDE in Fig.\ref{fig:clkk_main}, which turns out to be general for lensing shape constraint. Therefore, to simplify presentation, we only present the AdS-EDE result hereafter unless explicitly specified. The most constrained range is
roughly $80\lesssim L\lesssim400$, in accordance with the ACT+Planck bandpower
bins. The constrained bands in this region distribute
approximately evenly around $A(L)=1$, which requires that any
model describing the late Universe that is compatible with data to
predict lensing potential close to the concordance model in this
multiple range. This can also be seen from the theory results
(green) in Fig.\ref{fig:clkk_bk18}, while EDE and $\Lambda$CDM
constraints are generally different in the high and low $L$ tail
(mainly due to the significantly different prediction on $n_s$ in
EDE \cite{Ye:2021nej, Jiang:2022uyg, Jiang:2022qlj,
Jiang:2023bsz}), they overlap with each other in $80\lesssim L\lesssim400$. This
observation agrees with the finding of Ref.\cite{Planck:2018lbu, ACT:2023dou} that lensing reconstruction is compatible with $A_\text{lens}=1$ when marginalizing over a single
$A_\text{lens}$.


 \begin{figure}
     \centering
     \includegraphics[width=\linewidth]{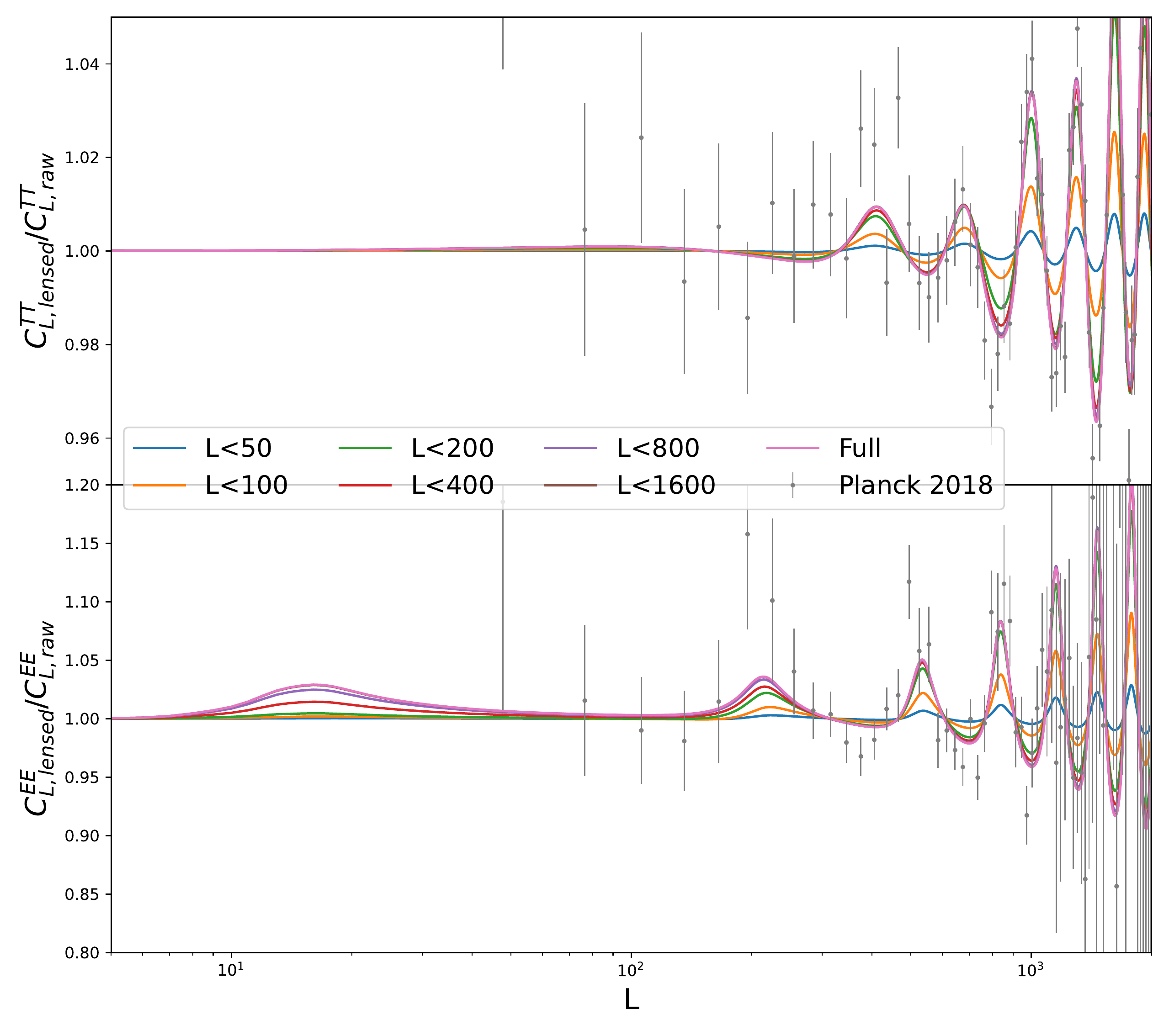}
     \caption{Comparison between the lensed and raw TT and EE spectra. Different line color indicates that the corresponding lensed spectra is produced with the $L<\#$ section of $C^{\phi\phi}_L$ (others set to zero). Points with error bars are binned power spectra of Planck 2018 \cite{Planck:2018vyg}.}
     \label{fig:lensed_cl}
 \end{figure}

 \begin{figure}
     \centering
     \includegraphics[width=0.9\linewidth]{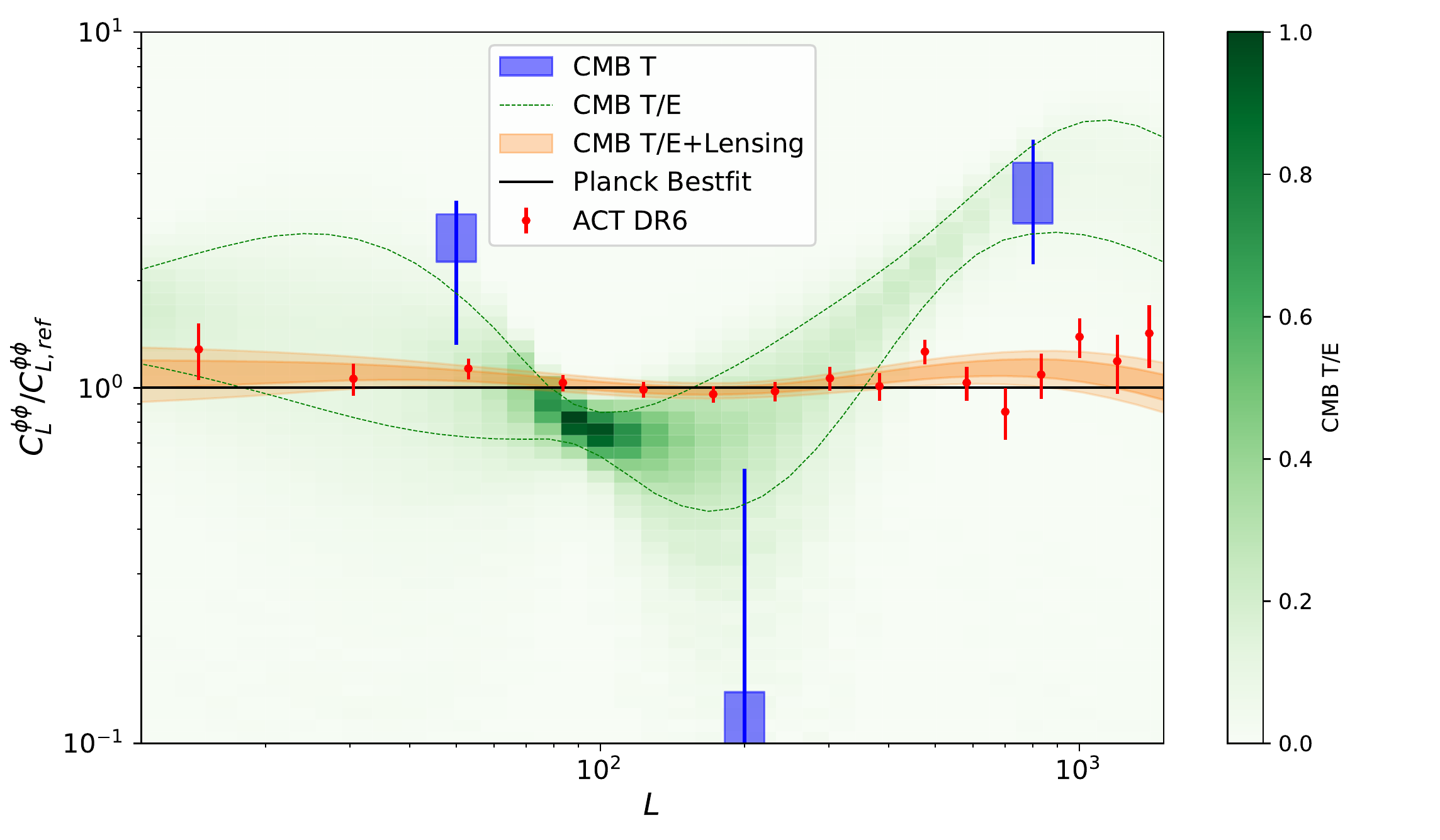}
     \caption{Comparison between the lensing shape constraints from CMB temperature (CMB T) and E polarization (CMB T/E) and CMB T/E+lensing reconstruction. For CMB T/E+Lensing, 68\% and 95\% posterior bands of $C^{\phi\phi}_L$ are plotted. For CMB T/E, color shading plots the posterior density distribution of the $C^{\phi\phi}_L$ values at 40 $L$'s log-evenly distributed in the range $10\le L\le1500$, with the corresponding 68\% C.L. band indicated by green dotted lines. For CMB T, the second node is unconstrained from below, causing the full shape constraint to be prior dominated. Therefore, we only plot the 68\% (blue rectangles) and 95\% (blue lines) constraints (upper bound for the second node) of $C^{\phi\phi}_L$ at $L=[50,200,800]$, corresponding to the positions of the three nodes $\{\ln \mathsf{A}_1, \ln \mathsf{A}_2, \ln
\mathsf{A}_3\}$.}
     \label{fig:clkk_cmbonly}
 \end{figure}

    \begin{figure}
    \includegraphics[width=0.9\linewidth]{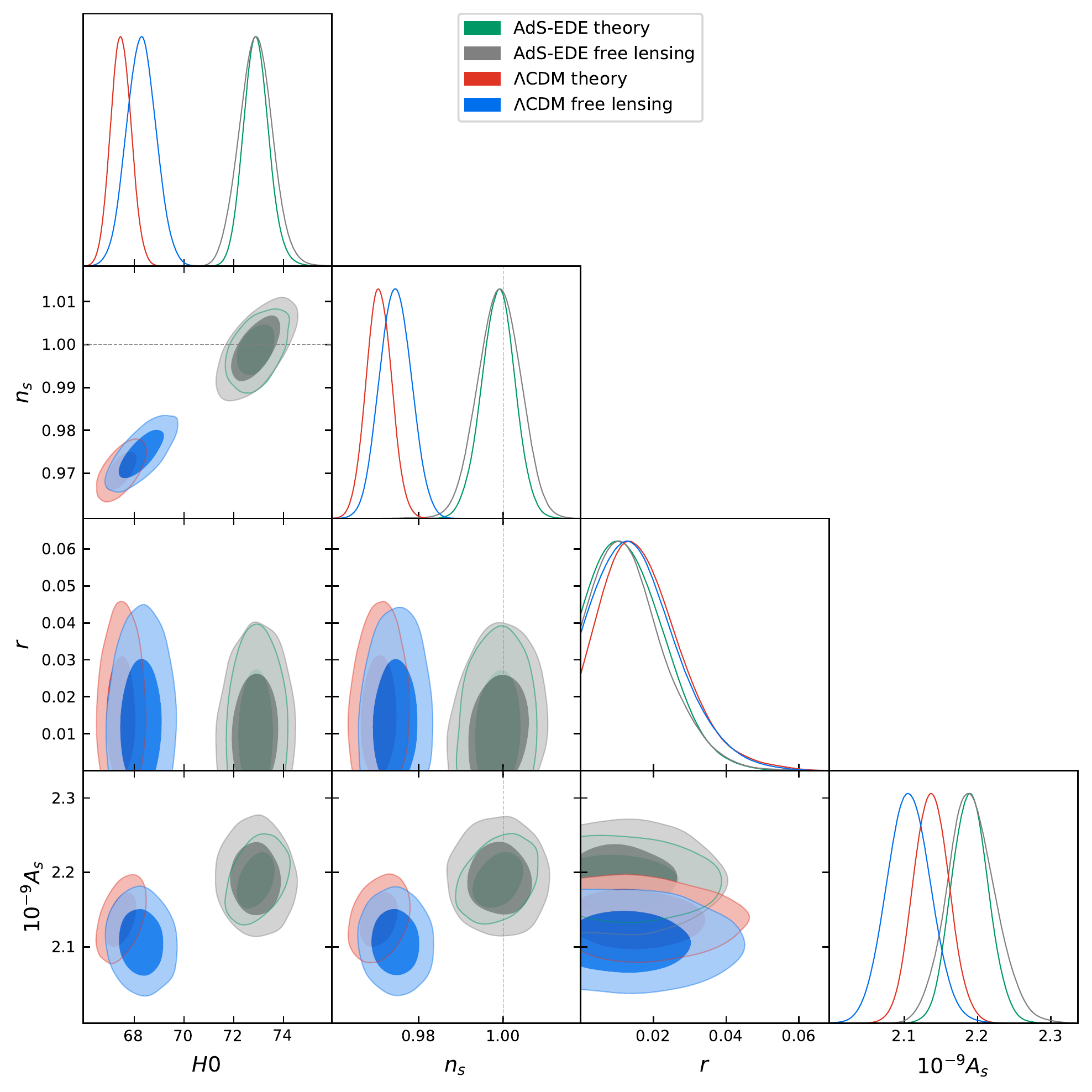}
        \caption{$68\%$ and $95\%$ posterior distributions of relevant cosmological parameters in the $\Lambda$CDM and AdS-EDE models with and without free lensing. The dataset used is baseline+BK18. Dotted lines mark the position of $n_s=1$.}
        \label{fig:rns_trig}
    \end{figure}

However, beyond the most constrained range $80\lesssim L\lesssim400$,
Fig.\ref{fig:clkk_main} indicates a preference, general to
$\Lambda$CDM and EDE, for enhanced lensing potential (compared
with the concordance model) in both the low and high $L$ tails. The enhancement in the low-$L$ tail can be at least partially attributed to the preference for additional lensing-like
smoothing in Planck data, sometimes referred to as the ``lensing anomaly".
Fig.\ref{fig:lensed_cl} shows how different sections of
$C^{\phi\phi}_L$ contribute to the lensed T and E spectra. The
oscillatory pattern in both spectra at high-$L$ is the
lensing smoothing effect, which according to
Ref.\cite{Planck:2018vyg}, accounts for most of the improvement in
fit for the lensing anomaly. This is also visible in Fig.\ref{fig:lensed_cl} where the Planck data points have slightly higher amplitude than the ``Full" line in the oscillatory region. Fig.\ref{fig:lensed_cl} shows that
$L<100$ has major contribution to smoothing while the effect of
$L>400$ is marginal. Fig.\ref{fig:clkk_cmbonly} further
compares the shape constraints from CMB temperature, polarization
and lensing reconstruction. Lensing reconstruction dominates the constraint in
the most constrained region $80\lesssim L\lesssim400$, while lensed temperature
and polarization spectra drive the preference for enhanced lensing
effect on both the low and high $L$ tails. E mode polarization
data reduces the enhancement at $L<100$, which is consistent with
the finding of Ref.\cite{Planck:2018vyg} that the lensing anomaly
becomes less obvious after inclusion of polarization spectra.

 In Fig.\ref{fig:lensed_cl}, the $L>400$ tail has significant contribution in the first two bumps of the E mode plot, especially the first bump is nearly exclusively contributed by $C^{\phi\phi}_L(L>200)$. This is due to the E to B mode conversion through CMB lensing and has already been noticed in Ref.\cite{Ye:2020btb}. It will thus be interesting to test if B-mode data provides additional constraints in the high-$L$ tail and, inversely, if the preferred enhancement of $C^{\phi\phi}_L$ at high-$L$ can change the derived bound on the tensor-to-scalar ratio $r$, defined as the ratio of primordial tensor power spectrum amplitude to scalar amplitude at $k=0.05 \mathrm{Mpc}^{-1}$. The enhancement at high $L$ of the baseline+BK18 result (red) compared with the baseline one (blue) in Fig.\ref{fig:clkk_bk18} shows that B-mode data indeed contributes non-trivially to the lensing shape constraint in $\Lambda$CDM. On the other hand, upper bounds on $r$ does not shift with the addition of free lensing, i.e. we obtain the same $r<0.037$ (95\% C.L.) in $\Lambda$CDM and $r<0.033$ (95\% C.L.) in AdS-EDE with and without free lensing, see Fig.\ref{fig:rns_trig} for relevant cosmological parameter constraints. We attribute the slightly different bounds than Ref.\cite{BICEP:2021xfz, Tristram:2021tvh, Ye:2022afu} to the different datasets used. Another interesting point in Fig.\ref{fig:rns_trig} and also Appendix-\ref{apdx:mcmc} is that the AdS-EDE cosmological constraints seems to be more ``robust" than $\Lambda$CDM against the addition of free lensing. This is because due to the nearly scale invariant $n_s$ \cite{Ye:2021nej, Jiang:2022qlj, Jiang:2022uyg, Jiang:2023bsz}, the theoretically predicted $C_{L,\text{th}}^{\phi\phi}$ in AdS-EDE is already close to the shape constraint at the high-$L$ tail, see Fig.\ref{fig:clkk_bk18}, while in $\Lambda$CDM, some additional parameter shifts are needed to accommodate the new lensing shape when free lensing is turned on. Interestingly, including BK18 on top of the baseline dataset in $\Lambda$CDM further increases the enhancement in $A(L)$ at high $L$.

    \section{Conclusion}\label{sec:conclusion}
In this paper, we studied how CMB data alone, including temperature
and polarization spectra as well as lensing reconstruction, can
constrain the overall shape of the lensing potential in
$\Lambda$CDM and EDE. We proposed a new function reconstruction method based on GP, which allowed us to marginalize over late-Universe effects such as lensing. As a consequence, the derived shape constraint does not rely on late Universe datasets
such as baryon acoustic oscillation, SNIa and galaxy surveys which
require assuming a specific cosmological model to bridge the
early and late Universe. On the other hand, our result does depend on the early Universe model assumed, e.g. EDE. The lensing reconstruction data has relatively weak dependence on models because the reconstruction procedure mainly relies on the observed spectrum. However, as shown in Fig.\ref{fig:clkk_cmbonly}, there is also part of the constraint coming from the smoothing of raw CMB T/E spectra. Both the smoothing effect and the raw spectra can depend on the model details of the prerecombination cosmology. It is quite interesting that AxiEDE and AdS-EDE produce very similar shape constraints in the end. We found current CMB lensing data
can constrain the shape of lensing on a broad multiple range and, according to Fig.\ref{fig:clkk_bk18}, the most stringent constraint is in $80\lesssim L\lesssim400$, where the cosmological models are required to produce $\Lambda$CDM-like lensing
potential. This is a strong constraint on the
possible behaviors of the late Universe during the period
relevant to CMB lensing (a broad redshift range around $z \sim 2$)
on these scales. On both the high and low-$L$ tails beyond this
range, the shape constraints are weaker but interestingly favor
enhanced lensing compared with $\Lambda$CDM prediction, which we
conclude is at least partially due to the preferred excess
lensing-like smoothing in Planck data \footnote{Our results only
indicate a similar possible preference in data. It does not
confirm the lensing anomaly in any statistical sense.}. We found
B-mode data can offer meaningful constraint on the lensing shape
at $L>400$ through its constraint on the lensing B-mode. Inclusion
of B-mode does not remove the preference for enhanced lensing at
$L>400$. Despite different shapes of theoretically predicted
lensing potentials, the lensing shape constraints from CMB data
are very similar in EDE and $\Lambda$CDM. However, we do notice
that, with free lensing, cosmological parameter constraints
experience nearly no change in AdS-EDE as opposed to $\Lambda$CDM
because the theory lensing potential shape in the former is closer
to the data preferred shape due to its very scale invariant $n_s$.

The constraints obtained in this paper can serve as a guideline
for model building. Given a late Universe physics model, a
necessary condition for such a model to be compatible with CMB is
that it should respect the most stringent constraints in
$80\lesssim L\lesssim400$ in Fig.\ref{fig:clkk_main},\ref{fig:clkk_bk18}. For
example, a direct observation from the constrained shape is that
any overall shift of the lensing potential amplitude will be difficult
to reconcile with CMB lensing, such as a constant $A_\text{lens}$
or positive spatial curvature \cite{Planck:2018vyg,
Handley:2019tkm, DiValentino:2019qzk, Vagnozzi:2020rcz}. This is in line with the observation that, despite the lensing anomaly, lensing reconstruction in both Planck 2018 \cite{Planck:2018vyg} and ACT DR6 \cite{ACT:2023dou} supports $A_{\rm{lens}}=1$. Given the
preference for enhanced lensing at $L>400$ and that galaxy weak
lensing surveys \cite{KiDS:2020suj, DES:2021wwk, Heymans:2020gsg}
seem to indicate reduced amplitude of matter power spectrum on
small scales, we are working on generalizing the
GP method presented here to the matter power spectrum and studying how
combining CMB lensing with galaxy weak lensing can further
constrain its shape.

\paragraph*{Acknowledgments}
We acknowledge the Xmaris cluster for providing computing
resources. Some figures are generated using GetDist
\cite{Lewis:2019xzd}. We thank Alessandra Silvstri for useful
comments and discussions. GY is supported by NWO and the Dutch
Ministry of Education, Culture and Science (OCW) (grant VI.Vidi.192.069). YSP is supported by NSFC, No.12075246
and the Fundamental Research Funds for the Central Universities.

    \appendix
    \section{More MCMC results}\label{apdx:mcmc}

 \begin{figure}
    \includegraphics[width=0.9\linewidth]{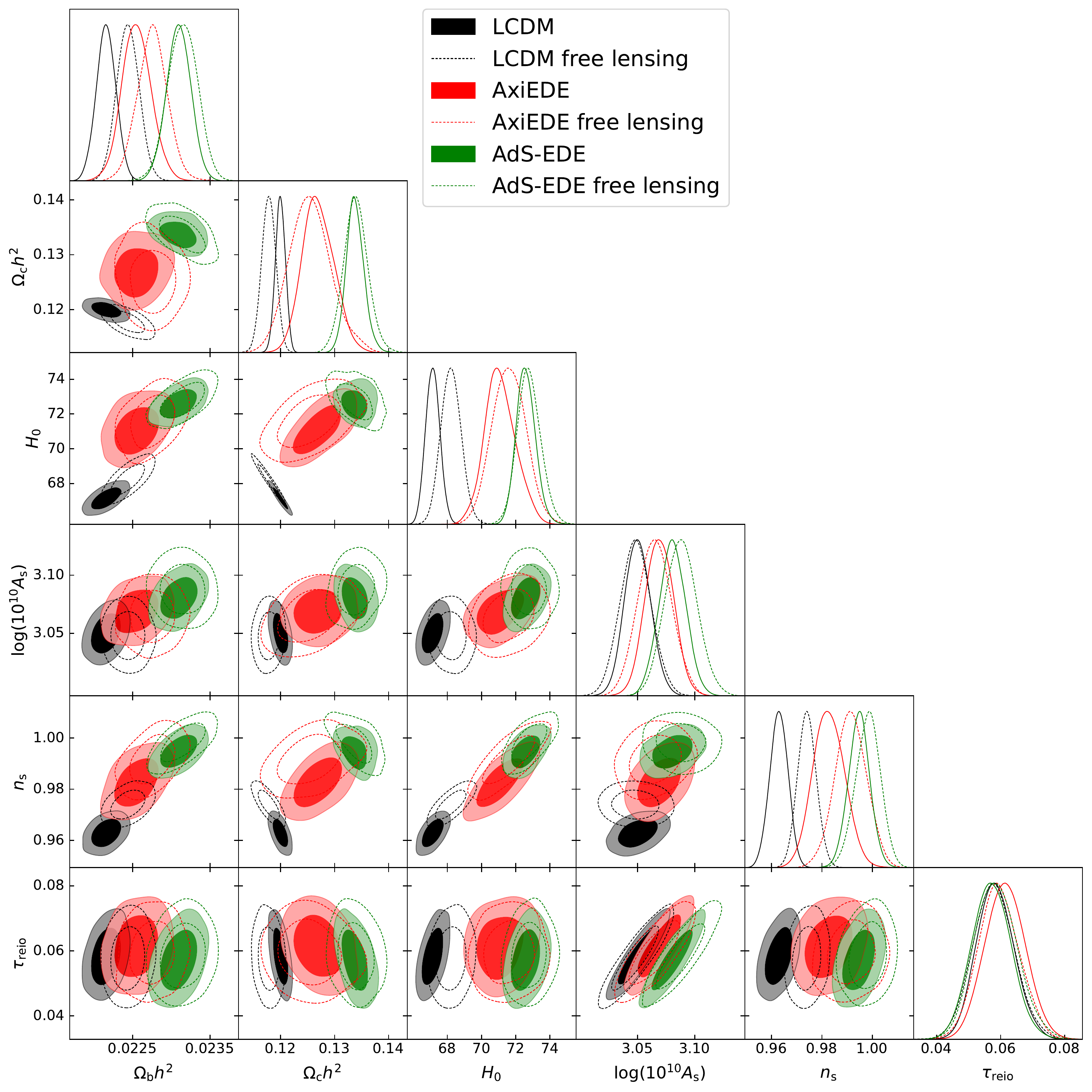}
        \caption{$68\%$ and $95\%$ posterior distributions of cosmological parameters in our baseline $\Lambda$CDM, AxiEDE and AdS-EDE models.}
        \label{fig:big_mcmc_trig}
    \end{figure}

\begin{table}
    \centering
    \begin{tabular}{|c|c||c|c|}
         \hline
          Parameter & Prior & Parameter & Prior \\
          \hline
          $\omega_b$ & Uniform$(-\infty,+\infty)$ & $f_{ede}$ & Uniform$(10^{-4},0.3)$\\
          \hline
          $\omega_{cdm}$ & Uniform$(-\infty,+\infty)$ & $\ln(1+z_c)$ & Uniform$(7.5,9.5)$\\
          \hline
          $H_0$ & Uniform$(-\infty,+\infty)$ & $\Theta_i$ & Uniform$(0, 3.1)$\\
          \hline
          $\ln10^{10}A_s$ & Uniform$(-\infty,+\infty)$ & $\ln A_{L_1}$ & Uniform$(-2,2)$\\
          \hline
          $n_s$ & Uniform$(-\infty,+\infty)$ & $\ln A_{L_2}$ & Uniform$(-2,2)$\\
          \hline
          $\tau_{reion}$ & Uniform$(0.004,+\infty)$ & $\ln A_{L_3}$ & Uniform$(-2,2)$\\
          \hline
    \end{tabular}
    \caption{Priors for all model parameters varied in our MCMC analysis.}
    \label{tab:mcmc_priors}
\end{table}

 \begin{table}
     \centering
     \footnotesize
     \begin{tabular}{|c|c|c|c|c|c|c|}
          \hline
          &\multicolumn{2}{c|}{$\Lambda$CDM}&\multicolumn{2}{c|}{AxiEDE}&\multicolumn{2}{c|}{AdS-EDE}\\
          \hline
          Parameters&theory&free lensing&theory&free lensing&theory&free lensing\\
          \hline
          $100\omega_b$&$2.216\pm 0.013        $&$2.245\pm 0.015            $&$2.253\pm 0.020        $&$2.276^{+0.019}_{-0.017}   $&$2.309\pm 0.016        $&$2.315\pm 0.019            $ \\
          \hline
          $\omega_{cdm}$&$0.1199\pm 0.0010        $&$0.1178\pm 0.0013          $&$0.1256^{+0.0039}_{-0.0032}$&$0.1253^{+0.0028}_{-0.0033}$&$0.1338^{+0.0015}_{-0.0017}$&$0.1342^{+0.0021}_{-0.0023}$  \\
          \hline
          $H_0$&$67.15\pm 0.43             $&$68.24\pm 0.59             $&$70.7\pm 1.1               $&$71.55\pm 0.83             $&$72.60^{+0.51}_{-0.59}     $&$72.79\pm 0.69             $  \\
          \hline
          $\ln10^{10}A_s$&$3.050\pm 0.012            $&$3.048\pm 0.014            $&$3.060^{+0.15}_{-0.18}      $&$3.064^{+0.013}_{-0.016}   $&$3.080\pm 0.012            $&$3.087\pm 0.014            $  \\
          \hline
          $n_s$&$0.9629\pm 0.0037$&$0.9740\pm 0.0038          $&$0.9793^{+0.0092}_{-0.0066}$&$0.9911^{+0.0061}_{-0.0053}$&$0.9947\pm 0.0043          $&$0.9981\pm 0.0049          $  \\
          \hline
          $\tau_{reion}$&$0.0583\pm 0.0058          $&$0.0577\pm 0.0066          $&$0.0614\pm 0.0064          $&$0.0595\pm 0.0070          $&$0.0572\pm 0.0061          $&$0.0583\pm 0.0065          $  \\
          \hline
          \hline
          $f_{ede}$&-&-&$0.078\pm 0.029            $&$0.088\pm 0.023            $&$0.1140^{+0.0031}_{-0.0078}$&$0.1154^{+0.0034}_{-0.0084}$  \\
          \hline
          $\ln(1+z_c)$&-&-&$8.289^{+0.345}_{-0.414}$&$8.312^{+0.143}_{-0.322}$&$8.142\pm 0.058            $&$8.120^{+0.054}_{-0.061}   $  \\
          \hline
          $\Theta_i$&-&-&$>2.54 \text{(95\%CL)}                  $&$2.79^{+0.19}_{-0.09}$&-&- \\
          \hline

     \end{tabular}
     \caption{Mean and 1$\sigma$ parameter constraints in our baseline $\Lambda$CDM, AxiEDE and AdS-EDE models.}
     \label{tab:big_mcmc_tab}
 \end{table}

 This appendix gathers some details of our baseline (i.e. CMB T/E+Lensing) MCMC analysis. Table.\ref{tab:mcmc_priors} lists the priors for all varied model parameters. Fig.\ref{fig:big_mcmc_trig} plots the posterior distributions in our baseline $\Lambda$CDM, AxiEDE and AdS-EDE analysis. The corresponding parameter constraints are reported in Table.\ref{tab:big_mcmc_tab}. Note that AxiEDE has a non-vanishing $f_{ede}$ here, which is due to the inclusion of an $H_0$ prior (R21) on top of the baseline datasets for this specific model, as explained in section-\ref{sec:results}. In both Fig.\ref{fig:rns_trig} and Fig.\ref{fig:big_mcmc_trig}, parameter posteriors with free lensing show some shift (not statistically significant) compared with the ones without. The shifts are generally smaller in Fig.\ref{fig:rns_trig} (baseline+BK18) than in Fig.\ref{fig:big_mcmc_trig} (baseline), which we attribute to the additional constraining power from BK18.

    \section{Impact of late ISW}\label{apdx:isw}

  \begin{figure}
    \includegraphics[width=0.9\linewidth]{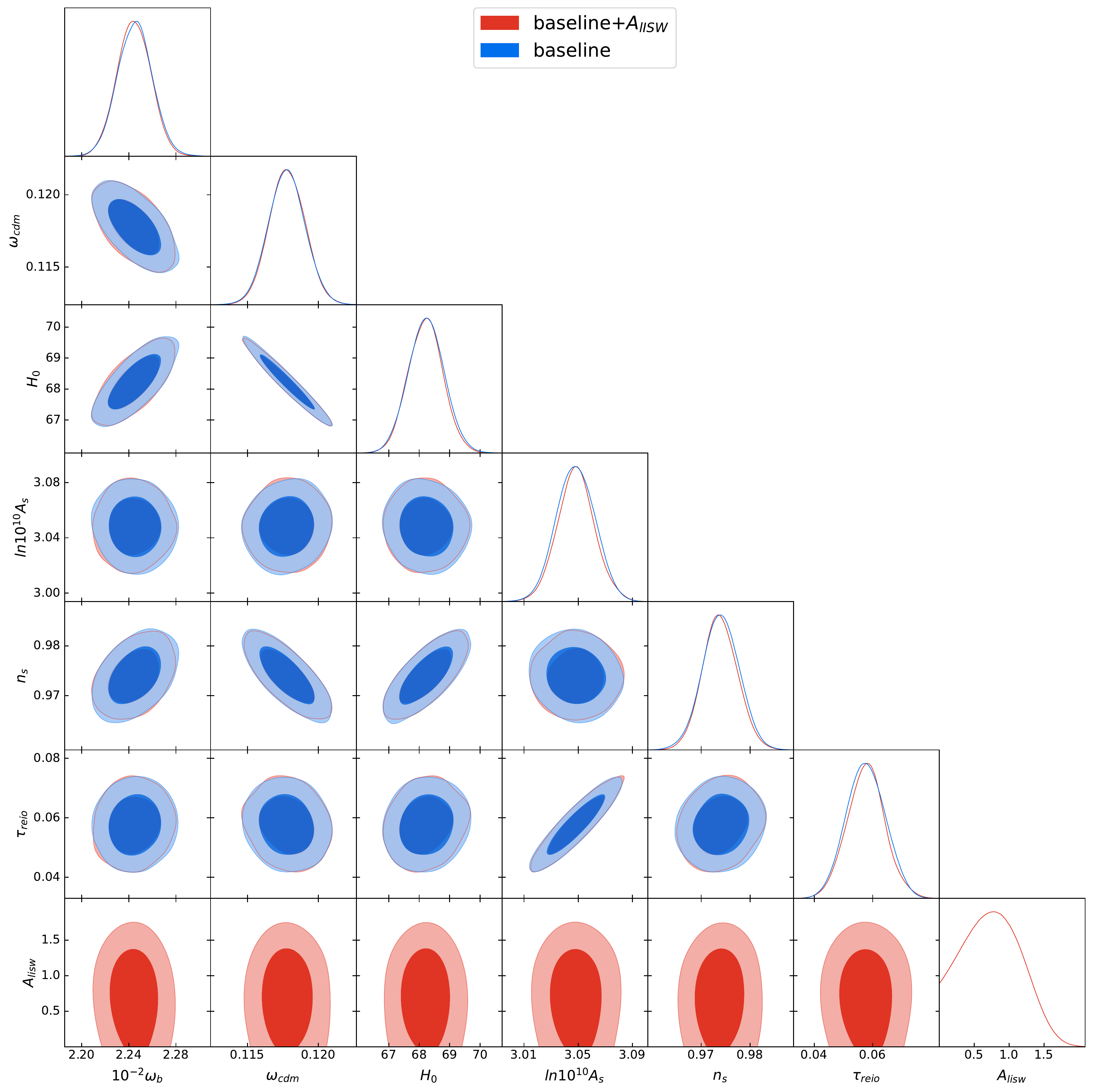}
        \caption{$68\%$ and $95\%$ posterior distributions of cosmological parameters in $\Lambda$CDM with $A_{LISW}=1$ and $A_{LISW}$ free.}
        \label{fig:lisw_trig}
    \end{figure}

 \begin{figure}
    \includegraphics[width=0.9\linewidth]{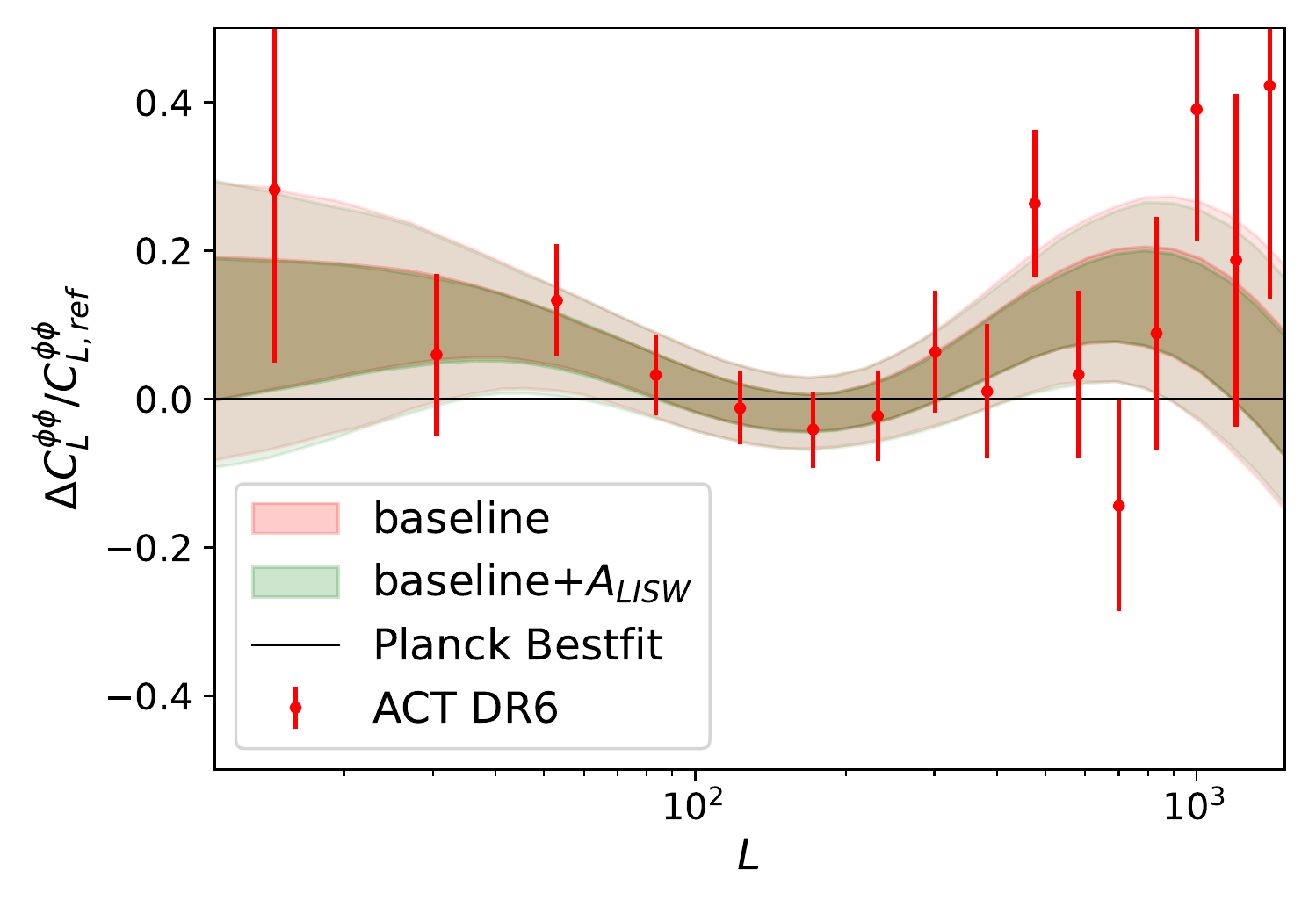}
        \caption{Lensing shape constraints in $\Lambda$CDM with $A_{LISW}=1$ and $A_{LISW}$ free.}
        \label{fig:lisw_clkk}
    \end{figure}

Another possible channel through which late Universe physics can
affect CMB observable is the late time ISW effect, mainly
originates from the dark energy dominated era. To test the impact
this effect, we rescale the late ISW ($z<120$) contribution to the
CMB power spectra by a free parameter $A_{LISW}$ and redo our
baseline analysis of $\Lambda$CDM. Fig.\ref{fig:lisw_trig}
compares the cosmological parameter constraints with $A_{LISW}=1$
and $A_{LISW}$ free, which shows that the constraint on $A_{LISW}$
is fairly weak and does not significantly impact parameter
constraints. Fig.\ref{fig:lisw_clkk} further illustrates that the
lensing shape constraints are also unaffected by fixing
$A_{LISW}=1$. Therefore, we fix $A_{LISW}=1$ in our baseline
analysis to reduce number of free parameters.

\section{Impact of correlation length $l$ and uncertainty $\sigma$}\label{apdx:correlation}

 \begin{figure}
    \includegraphics[width=0.9\linewidth]{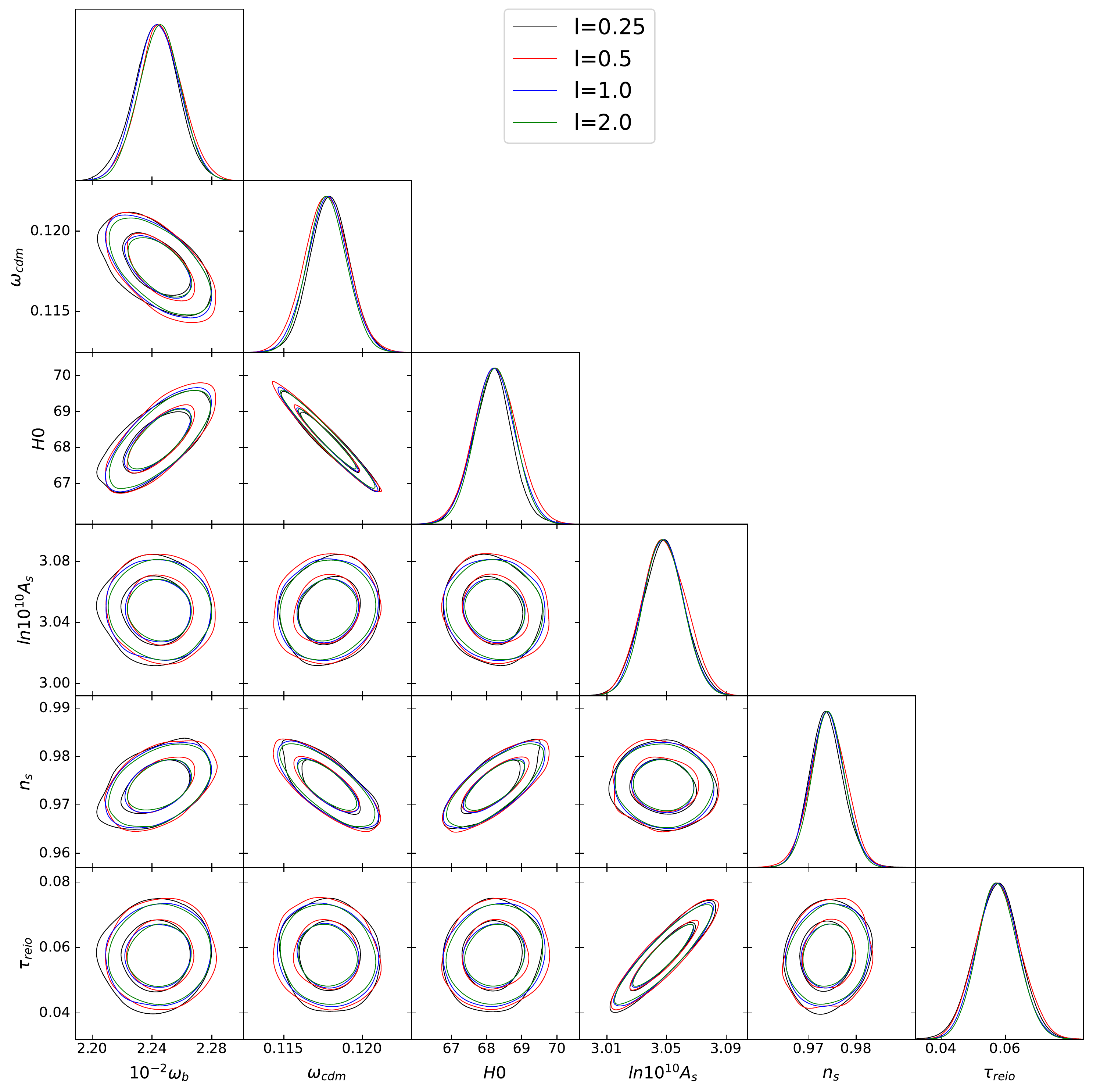}
        \caption{$68\%$ and $95\%$ posterior distributions of cosmological parameters in $\Lambda$CDM with different GP parameter $l$.}
        \label{fig:l_trig}
    \end{figure}

 \begin{figure}
    \includegraphics[width=0.48\linewidth]{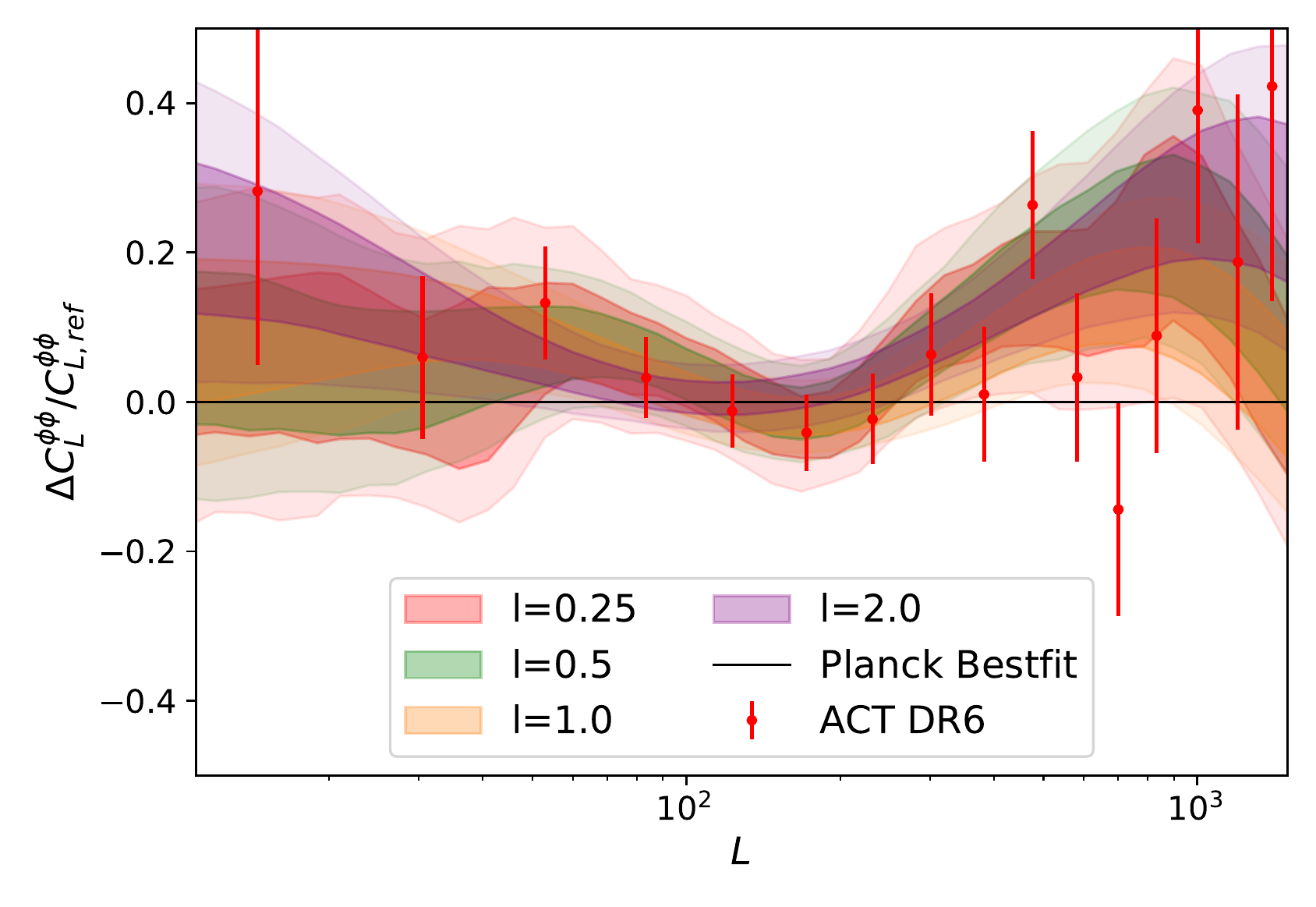}
    \includegraphics[width=0.48\linewidth]{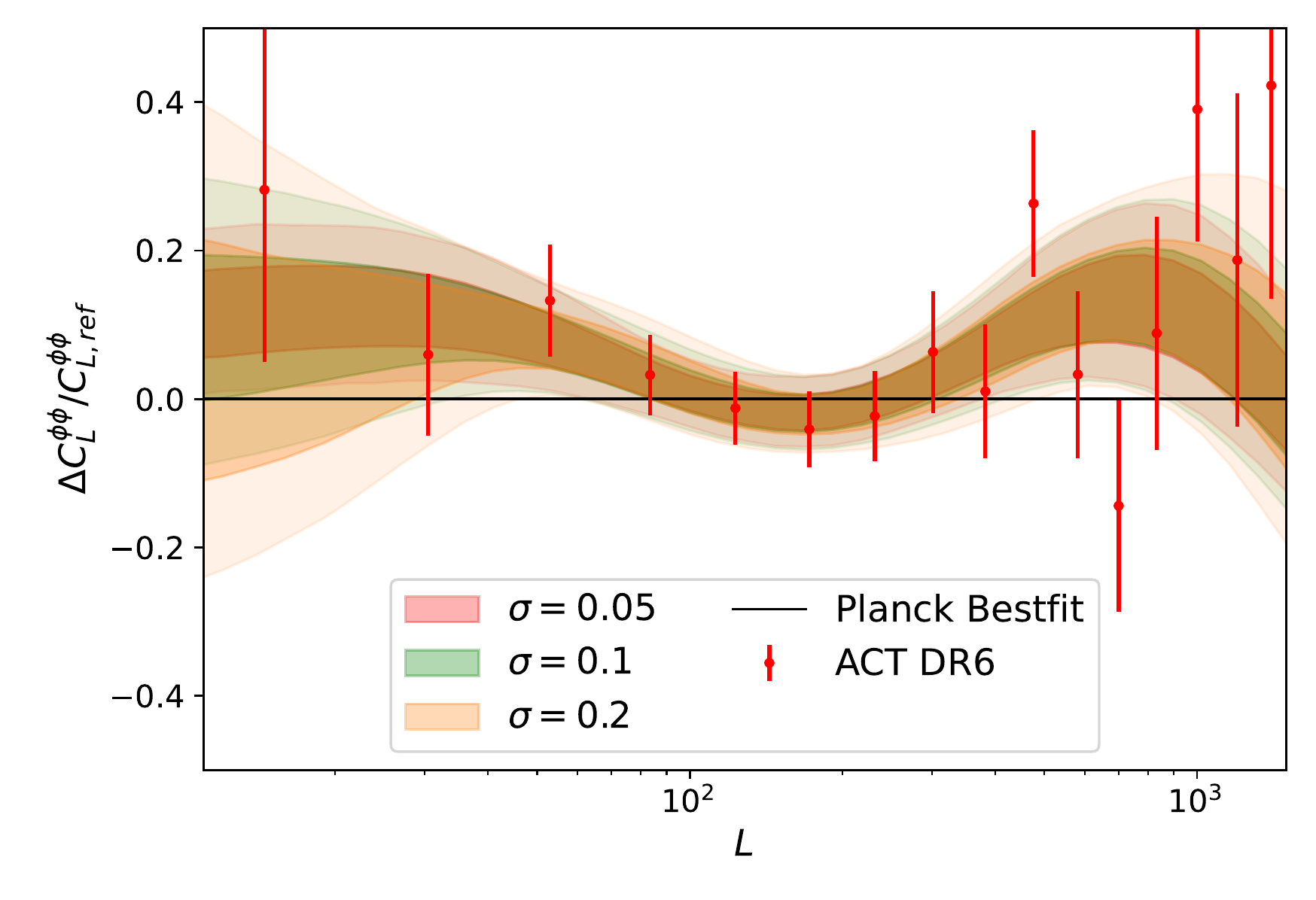}
        \caption{Lensing shape constraints in $\Lambda$CDM with different GP parameters. \textit{Left panel}: $l=\{0.25, 0.5,
1.0, 2.0\},~\sigma=0.1$. \textit{Right panel}: $l=1.0,~\sigma=\{0.05, 0.1, 0.2\}$.}
        \label{fig:gp_clkk}
    \end{figure}


In this section, we compare the impact of different values of the
GP parameters. Using the same model setups as
described in section-\ref{sec:results} but with $l=\{0.25, 0.5,
1.0, 2.0\},~\sigma=0.1$ and $l=1.0,~\sigma=\{0.05, 0.1, 0.2\}$, we redo our analysis against the baseline dataset. It
is clear Fig.\ref{fig:l_trig} that the choice of $l$ has
nearly no effects on the parameter constraints.
The left panel of Fig.\ref{fig:gp_clkk} shows that the general shape constraint is
also insensitive to the choice of $l$ in a wide multiple range. However,
too small $l$ means the GP will tend to sample more functions with
many random sharp features between two nodes and are not well
constrained by data. This in practice manifests as extremely low
acceptance rates and convergence speed in the MCMC analysis. We choose
to fix $l=1.0$ which is a good compromise between the acceptance
rate and good shape constraints. The right panel of Fig.\ref{fig:gp_clkk} shows that the main constraint at $L>40$ is also stable against a change in $\sigma$. We choose to fix $\sigma=0.1$ as it represents the typical relative error of the lensing reconstruction data.

    \bibliography{ref.bib}

\end{document}